\documentclass[journal]{IEEEtran}
\usepackage{amsmath}
\usepackage{amsfonts, amssymb}
\usepackage{algorithm}
\usepackage{algorithmic}
\usepackage{graphicx}
\usepackage{epstopdf}
\usepackage{subfigure}
\usepackage{booktabs}
\usepackage{threeparttable}
\usepackage{stfloats}
\usepackage{multirow}
\usepackage[noadjust]{cite}
\usepackage{tabularx}
\usepackage{amsmath} 
\usepackage{amssymb}  
\usepackage{hyperref}

\begin{document}

\title{Maximum Correntropy Adaptive Filtering Approach for Robust Compressive Sensing Reconstruction\thanks{Yicong He, Fei Wang and Badong Chen are with the Institute of Artificial Intelligence and Robotics, Xi'an Jiaotong University, China, e-mails: heyicong@stu.xjtu.edu.cn, wfx@xjtu.edu.cn, chenbd@xjtu.edu.cn.} \thanks{ Shiyuan Wang is with the School of Electronic and Information Engineering, Southwest University, China, e-mail: wsy@swu.edu.cn} \thanks{Jiuwen Cao is with the Key Lab for IOT and Information Fusion Technology of Zhejiang, Hangzhou Dianzi University, China, email: jwcao@hdu.edu.cn}}
\author{
Yicong He, Fei Wang, \IEEEmembership{Member,~IEEE}, Shiyuan Wang, \IEEEmembership{Member,~IEEE}, Jiuwen Cao, \IEEEmembership{Member,~IEEE}, Badong Chen, \IEEEmembership{Senior~Member,~IEEE}}
\IEEEpeerreviewmaketitle
\maketitle
\begin{abstract}
Robust compressive sensing(CS) reconstruction has become an attractive research topic in recent years. Robust CS aims to reconstruct the sparse signals under non-Gaussian(i.e. heavy tailed) noises where traditional CS reconstruction algorithms may perform very poorly due to utilizing $l_2$ norm of the residual vector in optimization. Most of existing robust CS reconstruction algorithms are based on greedy pursuit method or convex relaxation approach. Recently, the adaptive filtering framework has been introduced to deal with the CS reconstruction, which shows desirable performance in both efficiency and reconstruction performance under Gaussian noise. In this paper, we propose an adaptive filtering based robust CS reconstruction algorithm, called $l_0$ regularized maximum correntropy criterion($l_0$-MCC) algorithm, which combines the adaptive filtering framework and maximum correntropy criterion(MCC). MCC has recently been successfully used in adaptive filtering due to its robustness to impulsive non-Gaussian noises and low computational complexity. We analyze theoretically the stability of the proposed $l_0$-MCC algorithm. A mini-batch based $l_0$-MCC(MB-$l_0$-MCC) algorithm is further developed to speed up the convergence. Comparison with existing robust CS reconstruction algorithms is conducted via simulations, showing that the proposed $l_0$-MCC and MB-$l_0$-MCC can achieve significantly better performance than other algorithms.
\end{abstract}


%

\section{Introduction}
Compressive sensing(CS), or compressive sampling, is a novel sampling theory which demonstrates that a sparse or compressible signal can be well reconstructed with the sampling rate far less than Nyquist rate\cite{Donoho2006Compressed,candes2006compressive,Candes2006Robust,Candes2004Near,candes2008an}. CS takes advantages of the sparsity or compressibility of the signal, which is always available since many natural images or signals can be represented as the sparse signals in a certain domain\cite{Elad2006Image,Sch2006Efficient,Lustig2007Sparse,Potter2010Sparsity,Liu2016Object,Liu2015Robust}. An important procedure of the CS framework is to recover a sparse signal from insufficient number of measurements. In the last decade, many algorithms have been proposed to show accurate reconstruction performance, including greedy pursuit approaches such as matching pursuit(MP)\cite{Mallat1993Matching}, orthogonal matching pursuit(OMP)\cite{Tropp2007Signal}, iterative hard thresholding(IHT)\cite{Blumensath2008Iterative}, and convex relaxation approaches such as basis pursuit \cite{Chen2001Atomic}, iteratively reweighted least squares(IRLS)\cite{Chartrand2008Iteratively} and least absolute shrinkage and selection operator(LASSO)\cite{Tibshirani2011Regression}. Due to the low sampling rate as well as high reconstruction accuracy, CS has been widely used in many applications such as medical imaging\cite{Lustig2007Sparse} and radar imaging\cite{Potter2010Sparsity}.
\par Traditional CS algorithms often utilize $l_2$ norm of the residual vectors in optimization, which in general perform well under Gaussian noise assumption. However, in many real-world situations, due to impulsive disturbances or large outliers, the Gaussian assumption cannot be guaranteed, and the performance of traditional methods may degrade seriously. In recent years, robust CS reconstruction from noisy non-Gaussian(i.e. heavy-tailed) measurements draws much attention and many related algorithms have been proposed, such as $l_1$ based orthogonal matching pursuit($l_1$-OMP)\cite{Zeng2016Outlier}, Lorentzian-based iterative hard thresholding(LIHT)\cite{carrillo2013lorentzian} and Huber iterative hard thresholding(HIHT)\cite{Ollila2014Robust}. These algorithms utilize a certain robust objective function instead of the $l_2$ norm cost to solve the optimal solutions and the sparsity of the signal is supposed to know. However, the sparsity cannot always be available in practical applications. To solve this problem, several robust sparsity adaptive CS reconstruction algorithms are developed, such as $l_0$ regularized least absolute deviations($l_0$-LAD) solved by the coordinate descent approach\cite{Paredes2011Compressive} and $l_1$ regularized least absolute deviations($l_1$-LAD) solved by the alternating directions method of multipliers(ADMM)\cite{Yang2009Alternating,Xiao2013Primal}. A review on robust compressive sensing has been reported in\cite{Carrillo2016Robust}.
\par 
Recently, the adaptive filtering framework\cite{HaykinAdaptive,Sayed2003Fundamentals} has been introduced to CS reconstruction tasks\cite{Jin2010A}. The $l_0$ regularized least mean square($l_0$-LMS) algorithm, which is known to have good performance in sparse system identification, has been applied to sparse signal recovery. Measurement data are used recursively to ensure the convergence of the algorithm. The $l_0$-LMS algorithm demands less requirement in memory, and can achieve better reconstruction performance when dealing with large-scale CS reconstruction problem. To improve the convergence speed and reconstruction performance, the $l_0$-exponentially forgetting window LMS($l_0$-EFWLMS) was also proposed. Moreover, by using sign-error LMS instead of the traditional LMS algorithm, a robust gradient based CS reconstruction algorithm based on $l_0$-LAD regression was developed in\cite{ye2016robust}. Due to non-differentiability of the sign-error LMS, this algorithm is however mathematically intractable.
\par In recent years, a learning criterion in information theoretic learning(ITL)\cite{ITL} called Maximum Correntropy Criterion(MCC) has been successfully used in robust adaptive filtering\cite{Singh2009Using,Wang2015A,wu2015robust}. Correntropy is a nonlinear and local similarity measure. With a Gaussian kernel, correntropy involves all the even moments of the error and is insensitive to large outliers\cite{correntropy}. So far several adaptive algorithms under MCC have been proposed\cite{chen2015convergence, chen2012maximum,chen2014steady,Chen2015Maximum,Chen2016Generalized,Zhao2011Kernel}, which has shown desirable performance in robust system identification. Moreover, the Correntropy Matching Pursuit(CMP)\cite{Wang2016Correntropy} and Generalized Correntropy Matching Pursuit(GCMP)\cite{Loza2016Generalized} have been developed to deal with robust sparse representation. However, CMP and GCMP also assume that the sparsity is already known.
\par In this paper we propose an MCC based adaptive filtering framework to deal with robust CS reconstruction. By imposing an $l_0$ regularizer on MCC adaptive filtering algorithm and recursively using the data, the $l_0$-MCC algorithm for CS reconstruction is developed. The $l_0$-MCC algorithm takes advantages of both adaptive filtering framework and correntropy. Based on adaptive filtering framework, the proposed algorithm can adaptively estimate the sparsity and achieve signal reconstruction with high accuracy. Utilizing the MCC based objective function guarantees the robustness of the algorithm. The proposed $l_0$-MCC is mathematically tractable due to its simplicity and smoothness.
\par It is natural to follow the idea of $l_0$-EFWLMS and use more data within a 'window' at each iteration. However, in non-Gaussian noise environments, the bad affections of large outliers may sustain over iterations within the 'window'. Although correntropy can reduce the effects of outliers, persistent and periodic disturbances on estimation process may slow down the convergence speed and lead to unstable estimation results. To alleviate this deficiency, inspired by stochastic gradient descend(SGD) method, we relax the window to a subset, which randomly selects a set of training data chosen from all data with a fix number. This method is coincided with the mini-batch approach for SGD which aims to deal with the non-vanishing variance issue caused by SGD\cite{Konecny2014Mini}. With this inspiration, we propose a mini-batch based $l_0$-MCC(MB-$l_0$-MCC) to improve the convergence performance and robustness. Simulations are conducted to demonstrate the superior performance of $l_0$-MCC and MB-$l_0$-MCC in reconstruction accuracy and successful reconstruction probability compared with some existing robust CS reconstruction methods.
\par The remainder of the paper is organized as follows. In section II we briefly review the concepts of compressive sensing and correntropy. $l_0$-MCC and MB-$l_0$-MCC algorithms for CS reconstruction are then proposed in Section III. In Section IV, we analyze the stability of the $l_0$-MCC algorithm. Simulation results are presented in Section V, and finally conclusion is given in Section VI.
\section{Backgrounds}

\subsection{Compressive sensing}
Consider a real valued discrete signal $\boldsymbol{s} \in \mathbb{R}^{N \times 1}$. Suppose $\boldsymbol{s}$ is compressible, i.e. $\boldsymbol{s}$ can be represented as $\boldsymbol{s}=\boldsymbol{\Theta}\boldsymbol{x}$ where $\Theta\in \mathbb{R}^{N \times N}$ is a basis matrix and $\boldsymbol{x}$ is a sparse signal with sparsity $K\ll N$. In CS, the signal $\boldsymbol{s}$ is sampled(measured) by
\begin{equation}
\tilde {\boldsymbol{y}}=\boldsymbol{\Omega} \boldsymbol{s}=\boldsymbol{\Phi}\boldsymbol{x}
\end{equation}
where $\boldsymbol{\Omega}\in \mathbb{R}^{M\times N}$ is the measurement matrix, $\boldsymbol{\Phi}=\boldsymbol{\Omega}\boldsymbol{\Theta}$ is the sensing matrix and $\tilde {\boldsymbol{y}}\in \mathbb{R}^{M\times 1}$ is the measurement vector. Analysis shows that $\boldsymbol{\Phi}$ should obey the restricted isometry property(RIP)\cite{Cand2008The} to guarantee the recovery of sparse signal $\boldsymbol{x}$ with small number of measurements($M<N$), and it has also been proved that random matrix such as Gaussian random matrix and Rademacher random matrix can satisfy the RIP condition with a large probability.
\par In practice, due to perturbation of noise, the measurement vector $\tilde {\boldsymbol{y}}$ will be noisy, which can be described as
\begin{equation}
\label{CS1}
\boldsymbol{y}=\boldsymbol{\Phi} \boldsymbol{x}+\boldsymbol{v}
\end{equation}
where ${\boldsymbol{y}}$ is the noisy measurement and $\boldsymbol{v}$ is the additive noise vector. Thus the CS reconstruction problem can be expressed as solving an under-determined linear regression problem from a sensing matrix $\boldsymbol{\Phi}$ and a noisy measurement $\boldsymbol{y}$ with sparsity constraint.
\par In the context of adaptive filtering, the CS reconstruction problem in Eq.({\ref{CS1}}) can be solved by iteratively estimating the vector $\boldsymbol{x}$ from a finite data set $\boldsymbol{\Phi}$ and $\boldsymbol{y}$. In particular, the sensing matrix $\boldsymbol{\Phi}$ and measurement vector $\boldsymbol{y}$ can be rewritten as
$$\boldsymbol{\Phi}=[\boldsymbol{\phi}(1)~\boldsymbol{\phi}(2)~...~\boldsymbol{\phi}(M)]^T$$
$$\boldsymbol{y}=[y(1)~y(2)~...~y(M)]^T$$
thus each row of the sensing matrix $\boldsymbol{\Phi}$ and the corresponding element of $\boldsymbol{y}$ form the input and output data sequence $\{\boldsymbol{\phi}(i),y(i)\}_{i=1}^M$.
\par There are two main differences between CS reconstruction and traditional adaptive filtering. First, the number of training data in CS is limited, and the adaptive algorithm may not convergence to the steady state due to insufficient iteration numbers. To address this problem, the data of $\boldsymbol{\Phi}$ and $\boldsymbol{y}$ can be used recursively. Second, since the estimated vector $\boldsymbol{x}$ is sparse, one should add a regularizer such as the $l_0$ norm or $l_1$ norm to ensure the sparsity constraint. It has been shown that $l_0$ norm requires less number of measurements to recover $\boldsymbol{x}$ than $l_1$ norm\cite{Chartrand2007Exact}.

\subsection{Correntropy}
Correntropy is a local and nonlinear similarity measure between two random variables within a "window" in the joint space determined by the kernel width.
Given two random variables $X$ and $Y$, the correntropy is defined as
\begin{equation}
V(X, Y)= \boldsymbol{E}[\kappa (X, Y)]=\int \kappa_{\sigma} (x, y)dF_{XY}(x, y)
\end{equation}
where $\kappa _\sigma$ is a shift-invariant Mercer kernel, and $F _{XY}(x, y)$ denotes the joint distribution function of $(X, Y)$.
In practice, given a finite number of samples $ \{(x_i, y_i) \} _{i=1}^N$, the correntropy can be approximated by
\begin{equation}
\hat{V}(X, Y)= \frac{1}{N} \sum_{i=1}^N \kappa (x_i, y_i)
\end{equation}
\par
In general, the kernel function of correntropy is the Gaussian kernel given by
\begin{equation}
\label{Gkernel}
\kappa (x, y)= \frac{1}{\sqrt{2\pi}\sigma}\exp(-\frac{e^2}{2\sigma^2})
\end{equation}
where $e=x-y$ and $\sigma$ is the Gaussian kernel width.
\par Correntropy has been successfully used in adaptive filtering. By replacing the $l_2$ norm with correntropy, the optimal solution can be obtained by maximizing the following cost function
\begin{equation}
J_{MCC}= \frac{1}{N} \sum_{i=1}^N\exp(-\frac{e^2(i)}{2\sigma^2})
\end{equation}
which is called the maximum correntropy criterion(MCC). The MCC can also be formulated as minimizing the following C-loss function\cite{Singh2014The}
\begin{equation}
\label{C-loss}
J_{C-loss}= 1-\frac{1}{N} \sum_{i=1}^N\exp(-\frac{e^2(i)}{2\sigma^2})
\end{equation}
Without lose of generality, Eq.(\ref{C-loss}) is also referred to as the MCC cost function in this paper.

\section{MCC based algorithm for compressive sensing}

In this section we propose two methods to solve robust CS reconstruction problem under adaptive filtering framework. First, we introduce MCC based adaptive filtering to CS reconstruction problem and propose the $l_0$-MCC algorithm. Then, to speed up the convergence rate as well as maintain the robustness, the mini-batch based $l_0$-MCC(MB-$l_0$-MCC) algorithm is developed.

\subsection{\texorpdfstring{$l_0$}{l_0}-MCC algorithm}
In CS reconstruction problem, given the sensing matrix $\boldsymbol{\Phi}$ and measurement vector $\boldsymbol{y}$, the correntropy can be approximated as
\begin{equation}
\hat{V}_{CS}=\frac{1}{M}\sum_{i=1}^M\exp(-\frac{e^2(i)}{2\sigma^2})
\end{equation}
where $e(i)=y(i)-\boldsymbol{w}^T\boldsymbol{\phi}(i)$ is the predict residual at iteration $i$ and $\boldsymbol{w}$ is the estimation of $\boldsymbol{x}$. To ensure the sparsity of the estimated $\boldsymbol{w}$, by introducing the $l_0$ norm regularizer term, the optimal sparse solution can be obtained by minimizing the following cost function
\begin{equation}
\label{MCCCS}
J_{l_0-MCC-CS}(\boldsymbol{w})= 1-\frac{1}{M}\sum_{i=1}^M\exp(-\frac{e^2(i)}{2\sigma^2})+\lambda\|\boldsymbol{w}\|_0
\end{equation}
where $\lambda$ is the regularization parameter. Following the gradient descent method, by taking the derivative of Eq.(\ref{MCCCS}) and utilizing the instantaneous gradient value as the estimation of gradient, the weight update of $l_0$-MCC can be derived as
\begin{equation}
\begin{aligned}
\boldsymbol{w}(i+1)&=\boldsymbol{w}(i)+\mu \frac {\partial \hat J_{l_0-MCC-CS}(\boldsymbol{w})}{\partial \boldsymbol{w}(i)}\\
&=\boldsymbol{w}(i)+\mu\exp ( { - \frac{{{e^2}(i)}}{{2\sigma^2  }}} )e(i)\boldsymbol{\phi}(i)+\mu\lambda\nabla\|\boldsymbol{w}(i)\|_0
\end{aligned}
\end{equation}
where $\mu$ is the step size.
\par An important problem left is to calculate the derivation of $l_0$ term $\nabla\|\boldsymbol{w}(i)\|_0$. Since the $l_0$-norm is non-differentiable, a general way is to approximate $\nabla\|\boldsymbol{w}(i)\|_0$ by utilizing the following pairwise function
\begin{equation}
\begin{aligned}
\nabla\|\boldsymbol{w}(i)\|_0&\approx \boldsymbol{z}_{\beta}(\boldsymbol{w}(i))\\
&=[z_{\beta}(w_1(i)),z_{\beta}(w_2(i)),...,z_{\beta}(w_N(i))]^T
\end{aligned}
\end{equation}
where
\begin{equation}
z_{\beta}(w_m(i))=\left\{ {\begin{array}{*{20}{c}}
\beta^2w_m(i)+\beta~~ -1/\beta\leq w_m(i)<0\\
\beta^2w_m(i)-\beta~~~~~ 0<w_m(i)\leq 1/\beta
\end{array}} \right.
\end{equation}
$\boldsymbol{z}_{\beta}(\boldsymbol{w}(i))$ is called the zero attraction term. The parameter $\beta$ controls the attraction region and small coefficients within $[-1/\beta,1/\beta]$ will be attracted to zero gradually as iteration continues. Benefiting from the merit of most of the entries of $\boldsymbol{x}$ are zero, given a proper $\beta$, zero attraction can ensure the sparsity of the estimation as well as increase the convergence speed.

Therefore, the final weight update equation of $l_0$-MCC algorithm is
\begin{equation}
\label{l0MCC}
\boldsymbol{w}(i+1)=\boldsymbol{w}(i)+\mu\exp ( { - \frac{{{e^2}(i)}}{{2\sigma^2  }}} )e(i)\boldsymbol{\phi}(i)+\mu\lambda\boldsymbol{z}_{\beta}(\boldsymbol{w}(i))
\end{equation}
\par Note that the number of data may be insufficient, to ensure the sufficient iteration number for convergence, one should utilize the $\boldsymbol{\phi}(i)$ and $y(i)$ recursively\cite{Jin2010A}, i.e.
\begin{equation}
\boldsymbol{\phi}(i+M)=\boldsymbol{\phi}(i),y(i+M)=y(i).
\end{equation}
The pesudo code of $l_0$-MCC is summarized in Algorithm 1.
\begin{algorithm}
\caption{$l_0$-MCC Algorithm}
\begin{algorithmic}
 \STATE \emph{Initialization}
 \STATE choose step-size $\eta$, Gaussian kernel width $\sigma$, regularization parameter $\lambda$ and zero attraction parameter $\beta$\\initial iteration number $i=0$ and weight vector $\boldsymbol{w}(0)=\boldsymbol{0}$\\Set error tolerance $\varepsilon$ and maximum iteration number $C$
 \STATE \emph{Computation}
 \WHILE {$i<C$}
 \STATE \%extract the input vector and corresponding output vector from $\boldsymbol{\Phi}$ and $\boldsymbol{y}$
 \STATE $k=\mod(i,M)+1,\boldsymbol{x}(i)=\boldsymbol{\phi}(k),d(i)=y(k)$
 \STATE \%compute the output
 \STATE $a(i)=\boldsymbol{w}^T(i)\boldsymbol{x}(i)$
 \STATE \%compute the error
 \STATE $e(i)=d(i)-a(i)$
 \STATE \%update the weight vector based on gradient
 \STATE $ \boldsymbol{w}(i+1)=\boldsymbol{w}(i)+\mu\exp \left( { - \frac{{{e^2}(i)}}{{2\sigma^2  }}} \right)e(i)\boldsymbol{x}(i)$
 \STATE \%update the weight vector by zero attraction
 \STATE $\boldsymbol{w}(i+1)=\boldsymbol{w}(i+1) +\mu\lambda\boldsymbol{z}_{\beta}(\boldsymbol{w}(i))$
 \IF {$\|\boldsymbol{w}(i+1)-\boldsymbol{w}(i)\|^2<\varepsilon$}
 \STATE break
 \ENDIF
 \STATE \%update iteration number
 \STATE $i=i+1$
\ENDWHILE
\end{algorithmic}
\end{algorithm}
\par$Remark~1$: One can observe that when $\sigma\rightarrow\infty$ Eq.({\ref{l0MCC}}) will be equal to
\begin{equation}
\boldsymbol{w}(i+1)=\boldsymbol{w}(i)+\mu e(i)\boldsymbol{\phi}(i)+\mu\lambda\boldsymbol{z}_{\beta}(\boldsymbol{w}(i))
\end{equation}
which is the typical $l_0$-LMS algorithm for CS\cite{Jin2010A}. While when $\sigma$ is small, due to property of correntropy, the $l_0$-MCC algorithm can effectively reduce the bad effect when the residual $e(i)$ is large(caused by large outliers or impulsive noise). Moreover, since $\exp \left( { - \frac{{{e^2}(i)}}{{2\sigma^2  }}} \right)$ is associated with $\boldsymbol{w}(i)$, the convergence behaviour is no longer the same as $l_0$-LMS.

\subsection{Mini batch based $l_0$-MCC algorithm}
The $l_0$-MCC is simple and efficient. However, $l_0$-MCC uses only one pair of $\boldsymbol{\phi}(i)$ and $y(i)$ per iteration, thus the convergence will be slow. To improve the convergence speed, one would prefer more information in each iteration. In adaptive filtering, affine projection algorithm(APA)\cite{Member1984An} takes several most recent input vectors within a sliding window into consideration in each iteration and employs better convergence performance than instantaneous gradient based algorithm. Directly applying the MCC to APA yields the following cost function
\begin{equation}
\label{MCCAPA}
J_{MCC\!-\!APA}(\boldsymbol{w}(i))\!=\!\!\sum_{m=i-S+1}^i\!\!\exp\!\left(\!-\frac{\left(y(m)\!-\!\boldsymbol{w}^T(i)\boldsymbol{\phi}(m)\right)^2}{2\sigma^2}\right)
\end{equation}
where in each iteration the $S$ most recent input vectors and the corresponding observations are adopted for estimation.

The above cost function is based on tradition adaptive filtering method whose input vectors are assumed to be given in turn. However, in CS reconstruction problem, all the data are already known, therefore one can utilize data beyond traditional manner in non-Gaussian noise case. Inspired by SGD method, we again extend the cost function in Eq.({\ref{MCCAPA}}) to
\begin{equation}
\label{MBMCC}
J_{MB\!-\!MCC}(\boldsymbol{w}(i))\!=\! \sum_{k=1}^S \exp \left( { \!- \frac{{{\left(y(r(k))\!-\!\boldsymbol{w}^T(i)\boldsymbol{\phi}(r(k))\right)^2}}}{{2\sigma^2  }}} \right)
\end{equation}
where $\boldsymbol{r}=[r(1)~r(2)~...~r(S)]^T\in\mathbb{N}_{+}^{S\times1}$ is the index vector whose elements are uniformly and randomly chosen from $[1,M]$. 
\par The new cost function in Eq.({\ref{MBMCC}}) is coincide with the mini-batch method used in SGD method with a different initialization. Indeed the cost inherits the good properties of mini-batch method for SGD, such as better stochastic estimate of the gradient and faster convergence rate, this method is more suitable for non-Gaussian noise. Particularly, in non-Gaussian CS reconstruction problem, if the sliding window method is directly used in Eq.({\ref{MCCAPA}}), a large outlier may cause persistent bad disturbance as long as $S$ iterations, and periodic disturbance is imposed due to recursive usage of data. Moreover, if the measurement vector $\boldsymbol{y}$ contains too many large outliers, the performance will become worse. In contrast, random selection of training data at each iteration can avoid persistent and periodical bad influence of large outliers as well as increase the convergence speed, resulting more robust and accurate performance for sparse signal reconstruction in non-Gaussian noise environments.
\par Similar to $l_0$-MCC, by adding the $l_0$ norm to Eq.({\ref{MBMCC}}), a new algorithm called the mini-batch based $l_0$-MCC(MB-$l_0$-MCC) algorithm is proposed, where the estimation can be obtained by the following weight update equation
%
\begin{equation}
\boldsymbol{w}(i)=\boldsymbol{w}(i-1)+\mu \boldsymbol{X}^T(i)\boldsymbol{G}\boldsymbol{e}(i)
\end{equation}
where
\begin{equation}
\begin{aligned}
&\boldsymbol{X}(i)=[\boldsymbol{\phi}(r(1)),\boldsymbol{\phi}(r(2)),...,\boldsymbol{\phi}(r(S))]^T\\
&\boldsymbol{d}(i)=[y(r(1)),y(r(2)),...,y(r(S))]^T\\
&\boldsymbol{e}(i)=\boldsymbol{d}(i)-\boldsymbol{X}^T(i)\boldsymbol{w}(i)
\end{aligned}
\end{equation}
and $\boldsymbol{G}\in \mathbb{R}^{S\times S}$ is a diagonal matrix with entries
\begin{equation}
\boldsymbol{G}(i,i)=\exp \left( { - \frac{{{e^2}(i)}}{{2\sigma^2  }}} \right), i=1,...,S
\end{equation}
The pesudo code of MB-$l_0$-MCC is summarized in Algorithm 2.
\begin{algorithm}
\caption{MB-$l_0$--MCC Algorithm}
\begin{algorithmic}
 \STATE \emph{Initialization}
 \STATE choose step-size $\eta$, Gaussian kernel width $\sigma$, batch size $S$, regularization parameter $\lambda$ and zero attraction parameter $\beta$\\initial iteration number $i=0$ and weight vector $\boldsymbol{w}(0)=\boldsymbol{0}$\\Set error tolerance $\varepsilon$ and maximum iteration number $C$
 \STATE \emph{Computation}
 \WHILE {$i<C$}
 \STATE \%randomly select $S$ input vector and corresponding output vector from $\boldsymbol{\Phi}$ and $\boldsymbol{y}$
 \STATE $\boldsymbol{r}=randselect(S,M),\boldsymbol{X}(i)=\boldsymbol{\phi}(\boldsymbol{r}),\boldsymbol{d}(i)=y(\boldsymbol{r})$
 \STATE \%compute the output vector
 \STATE $\boldsymbol{a}(i)=\boldsymbol{X}^T(i)\boldsymbol{w}(i)$
 \STATE \%compute the error vector
 \STATE $\boldsymbol{e}(i)=\boldsymbol{d}(i)-\boldsymbol{a}(i)$
 \STATE \%update the weight vector based on gradient
 \STATE $ \boldsymbol{w}(i+1)=\boldsymbol{w}(i)+\mu \boldsymbol{X}^T(i)\boldsymbol{G}\boldsymbol{e}(i)$
 \STATE \%update the weight vector by zero attraction
 \STATE $\boldsymbol{w}(i+1)=\boldsymbol{w}(i+1) +\mu\lambda\boldsymbol{z}_{\beta}(\boldsymbol{w}(i))$
 \IF {$\|\boldsymbol{w}(i+1)-\boldsymbol{w}(i)\|^2<\varepsilon$}
 \STATE break
 \ENDIF
 \STATE \%update iteration number
 \STATE $i=i+1$
\ENDWHILE
\end{algorithmic}
\end{algorithm}

\section{Stability analysis of \texorpdfstring{$l_0$}{l_0}-MCC}
In this section, we analysis the stability of the proposed $l_0$-MCC algorithm. We first derive the condition on step size $\mu$ for mean square stability and then analysis the choice of $\mu$ under some specific conditions.
\par For tractability, in the following analysis we assume that the non-Gaussian noise $v(i)$ has finite variance. This assumption is widely used in analysis, and is always satisfied in practice. Moreover, the performance under infinite variance noise(i.e. $\alpha$-stable noise) case will be illustrated by simulation. Without loss of generality, during the analysis we also assume that each entry of $\boldsymbol{\Phi}$ is i.i.d random variable with variance $\sigma_a^2$.

\subsection{Conditions for mean-square stability}
From Eq.({\ref{CS1}}) one can obtain
\begin{equation}
y(i)=\boldsymbol{x}^T\boldsymbol{\phi}(i)+v(i)
\end{equation}
So the residual error $e(i)$ can be derived as
\begin{equation}
\label{errorresidual}
e(i)=y(i)-\boldsymbol{w}(i)^T\boldsymbol{\phi}(i)=\boldsymbol{\widetilde{w}}(i)^T\boldsymbol{\phi}(i)+v(i)
\end{equation}
where $\boldsymbol{\widetilde{w}}(i)=\boldsymbol{x}-\boldsymbol{w}(i)$ is the weight error vector at iteration time $i$.
Substituting Eq.({\ref{errorresidual}}) into Eq.({\ref{l0MCC}}) and subtracting both sides from $\boldsymbol{x}$ yield
\begin{equation}
\begin{aligned}
\label{errorresidual2}
\boldsymbol{\widetilde{w}}(i+1)&\!=\!\boldsymbol{\widetilde{w}}(i)-\mu\exp \left( { - \frac{{{e^2}(i)}}{{2\sigma^2  }}} \right)e(i)\boldsymbol{\phi}(i) -\mu\lambda\boldsymbol{z}_{\beta}(\boldsymbol{w}(i))\\
&\!=\! [\boldsymbol{I}\!-\!\mu G(e(i))\boldsymbol{\phi}(i)\boldsymbol{\phi}^T(i)]\boldsymbol{\widetilde{w}}(i)\!-\!\mu G(e(i))v(i)\boldsymbol{\phi}(i)\\
&~~~-\mu\lambda\boldsymbol{z}_{\beta}(\boldsymbol{w}(i))
\end{aligned}
\end{equation}
where we use $G(e(i))=\exp \left( { - \frac{{{e^2}(i)}}{{2\sigma^2  }}} \right)$ for simplicity.
Postmultipling both sides of Eq.({\ref{errorresidual2}}) with their respective transposes,
then taking the trace operation, and finally taking the expectation can we obtain the following mean square recursion
\begin{equation}
\begin{aligned}
\label{MSR}
E[\|\boldsymbol{\widetilde{w}}(i+1)\|^2]&=E[\|\boldsymbol{\widetilde{w}}(i)\|^2]-2\mu H(i)-2\mu J(i)\\
&~~~+ \mu^2 K(i)-2\mu\lambda M(i)+ \mu\lambda^2 Q(i)\\
&~~~ + \mu^2 T(i) +2\mu^2 Y(i) + 2\mu^2\lambda P(i)
\end{aligned}
\end{equation}
where
\begin{equation}
\begin{aligned}
\label{Bi}
H(i)&=E[G(e(i))(\boldsymbol{\phi}^T(i)\boldsymbol{\widetilde{w}}(i))^2]\\
J(i)&=E[G(e(i))v(i)\boldsymbol{\phi}^T(i)\boldsymbol{\widetilde{w}}(i)]\\
K(i)&=E[G^2(e(i))\|\boldsymbol{\phi}(i)\|^2(\boldsymbol{\phi}^T(i)\boldsymbol{\widetilde{w}}(i))^2]\\
M(i)&=E[\boldsymbol{\widetilde{w}}^T(i)\boldsymbol{z}_{\beta}(\boldsymbol{w}(i))]\\
Q(i)&=E[\boldsymbol{z}^T_{\beta}(\boldsymbol{w}(i))\boldsymbol{z}_{\beta}(\boldsymbol{w}(i))]\\
T(i)&=E[G^2(e(i))v^2(i)\|\boldsymbol{\phi}(i)\|^2]\\
Y(i)&=E[G^2(e(i))v(i)\|\boldsymbol{\phi}(i)\|^2(\boldsymbol{\phi}^T(i)\boldsymbol{\widetilde{w}}(i))]\\
P(i)&=E[G(e(i))e(i)(\boldsymbol{\phi}^T(i)\boldsymbol{z}_{\beta}(\boldsymbol{w}(i)))]
\end{aligned}
\end{equation}
and $E[\|\boldsymbol{\widetilde{w}}(i)\|^2]$ is denoted as mean square derivation(MSD) at iteration $i$. Note that $H(i)$ and $K(i)$ contains second order of $\boldsymbol{\widetilde{w}}(i)$ and may contain the term $\|\boldsymbol{\widetilde{w}}(i)\|^2$. For further analysis, we assume that $H(i)$ and $K(i)$ are represented as
\begin{equation}
\begin{aligned}
\label{KH}
K(i)=C_K(i)E[\|\boldsymbol{\widetilde{w}}(i)\|^2]\\
H(i)=C_H(i)E[\|\boldsymbol{\widetilde{w}}(i)\|^2]
\end{aligned}
\end{equation}
where $C_K(i)$ and $C_H(i)$ are coefficients at the $i$-th iteration.
Thus, substituting Eq.({\ref{KH}}) to Eq.({\ref{MSR}}) we can obtain
\begin{equation}
\begin{aligned}
\label{MSR2}
\!E[\|\boldsymbol{\widetilde{w}}(i+1)\|^2]\!&=\!\left(1\!-\!2\mu C_{\!H\!}(i)\!+\!\mu^2C_{\!K\!}(i)\right)\!E[\|\boldsymbol{\widetilde{w}}(i)\|^2]\!+\!B(i)\\
\end{aligned}
\end{equation}
where
\begin{equation}
\begin{aligned}
\label{Bii}
B(i)=&-2\mu J(i)-2\mu\lambda M(i)+ \mu\lambda^2 Q(i)\\
&+ \mu^2 T(i) +2\mu^2 Y(i) + 2\mu^2\lambda P(i)
\end{aligned}
\end{equation}
$B(i)$ can be proved to be bounded(the proof is given in Appendix A). To guarantee the convergence to steady state in the mean square sense, the
magnitude of coefficient of $E[\|\boldsymbol{\widetilde{w}}(i)\|^2]$ must be less than unity, leading to the following condition
\begin{equation}
|1-2\mu C_H(i)+\mu^2C_K(i)|<1
\end{equation}
Thus the step size $\mu(i)$ at time $i$ should satisfy the following inequations to guarantee the convergence of Eq.({\ref{MSR2}})
\begin{equation}
\label{stepsize}
\left\{
\begin{aligned}
&\mu(i)<\frac{2C_H(i)}{C_K(i)} \\
&\mu^2(i)C_K(i)-2\mu(i) C_H(i)+2>0 \\
\end{aligned}
\right.
\end{equation}
and then $E[\|\boldsymbol{\widetilde{w}}(i)\|^2]$ will finally converge to the steady state with the MSD
\begin{equation}
E\left[ {{{\left\| {\tilde {\boldsymbol{w}}(\infty)} \right\|}^2}} \right]=\frac{B(\infty)}{2\mu C_H(\infty)-\mu^2 C_K(\infty)}
\end{equation}
where $\infty$ denotes $i\rightarrow\infty$.
\par The above theoretical analysis is generally difficult to be applied in practice. Specifically, since the exact values of $C_H(i)$ and $C_K(i)$ are hard to obtain, the condition on step size $\mu$ at each ietration is difficult to calculate. In the following we derive some sufficient conditions on $\mu$ under some specific assumptions, which can be used in practice.

\subsection{Sufficient condition for convergence under Rademacher sensing matrix}
Rademacher matrix is a commonly used sensing matrix, with the entries independently generated from ${-\sigma_a}$ and ${\sigma_a}$ with equal probability.
In this part we derive the sufficient condition of $\mu$ using Rademacher sensing matrix. It is obviously that both $C_H(i)$ and $C_K(i)$ are non-negative values. Moreover, one can obtain from Eq.({\ref{Bi}}) that
\begin{equation}
\begin{aligned}
H(i)&=E[G(e(i))(\boldsymbol{\phi}^T(i)\boldsymbol{\widetilde{w}}(i))^2]\\
&\leq E[(\boldsymbol{\phi}^T(i)\boldsymbol{\widetilde{w}}(i))^2]\\
&=\sigma_a^2E[\|\boldsymbol{\widetilde{w}}(i)\|^2]
\end{aligned}
\end{equation}
Thus $C_H(i)$ has the range
\begin{equation}
\label{CHi}
0\leq C_H(i)\leq\sigma_a^2
\end{equation}
Also, from Eq.({\ref{Bi}}) one can obtain
\begin{equation}
\begin{aligned}
\label{KiR}
K(i)&=E[G^2(e(i))\|\boldsymbol{\phi}(i)\|^2(\boldsymbol{\phi}^T(i)\boldsymbol{\widetilde{w}}(i))^2]\\
&\leq N\sigma_a^2E[G(e(i))(\boldsymbol{\phi}^T(i)\boldsymbol{\widetilde{w}}(i))^2]\\
&= N\sigma_a^2C_H(i)E[\|\boldsymbol{\widetilde{w}}(i)\|^2]
\end{aligned}
\end{equation}
where the fact $\|\boldsymbol{\phi}(i)\|^2=N\sigma_a^2$ and $G(e(i))\in(0,1]$ are adopted to get Eq.(\ref{KiR}). Thus
\begin{equation}
\label{CKi}
0\leq C_K(i)\leq N\sigma_a^2C_H(i)
\end{equation}
Substituting Eq.({\ref{CHi}}) and Eq.({\ref{CKi}}) into Eq.({\ref{stepsize}}) yields the following sufficient condition on step size
\begin{equation}
\label{muRade}
0<\mu<\frac{2}{N\sigma_a^2}
\end{equation}
Therefore, given a N-dimensional signal, and the sensing matrix with variance $\sigma_a^2$, one can get an explicit sufficient condition of $\mu$ to guarantee the convergence.
\par $Remark~3$:The analysis above is also suitable for many sensing matrix such as sparse matrix\cite{Gilbert2008Sparse} and partial Hadamard matrix\cite{Pinkus1985n,Tsaig2006Extensions}, whose entries are bounded. But Eq.(\ref{KiR}) is no longer hold for Gaussian random sensing matrix. Fortunately, in the following we will show that due to use of Gaussian kernel in correntropy, one can still get the explicit result for $\mu$ under some specific conditions.


\subsection{Sufficient condition for convergence under Gaussian sensing matrix: bounded noise case}
In this part we will derive the sufficient condition under bounded noise with Gaussian random sensing matrix $\boldsymbol{\Phi}$. The entries of $\boldsymbol{\Phi}$ are assumed to be i.i.d Gaussian distribution with variance $\sigma_a^2$. Using integral method to calculate the expectation we can obtain
\begin{equation}
\begin{aligned}
\label{KHE}
H(i)&=E[G(e(i))(\boldsymbol{\phi}^T(i)\boldsymbol{\widetilde{w}}(i))^2]\\
&=E\left[ P_H(i){\left\| \boldsymbol{\widetilde{w}}(i) \right\|^2}\right]\\
K(i)&=E[G^2(e(i))\|\boldsymbol{\phi}(i)\|^2(\boldsymbol{\phi}^T(i)\boldsymbol{\widetilde{w}}(i))^2]\\
&\leq E[G(e(i))\|\boldsymbol{\phi}(i)\|^2(\boldsymbol{\phi}^T(i)\boldsymbol{\widetilde{w}}(i))^2]\\
&=E\left[P_K(i){\left\| \boldsymbol{\widetilde{w}}(i) \right\|^2}\right]
\end{aligned}
\end{equation}
where
\begin{equation}
\begin{aligned}
\label{PHiKi}
P_H(i)&={\frac{{\sigma}}{{{p(i)^{\frac{1}{2}}}}}}{\frac{{q(i)}}{{{p(i)^2}}}} \sigma _a^2 \exp \left( { - \frac{{{v^2(i)}}}{{2p(i)}}} \right)\\
P_K(i)&\!=\!{\frac{{\sigma}}{{{p(i)^{\frac{1}{2}}}}}}\!\!\left[ \!{\frac{{\left( {N \!-\! 1} \right)q(i)}}{{{p(i)^2}}} \!+\! \frac{{2{\sigma ^2}\!\!\left( {q(i) \!+\! \sigma _a^2{\left\| \boldsymbol{\widetilde{w}}(i) \right\|^2}{v^2(i)}} \right)\!}}{{{p(i)^3}}}} \right.\\
&~~~~~~~~~~~~\left.{\!+ \frac{{{q(i)^2}}}{{{p(i)^4}}}} \right]\!\sigma _a^4\exp \left( { - \frac{{{v^2(i)}}}{{2p(i)}}} \right)
\end{aligned}
\end{equation}
with
\begin{equation}
\begin{aligned}
p(i) &= \sigma^2 + \sigma _a^2{\left\| \boldsymbol{\widetilde{w}}(i) \right\|^2}\\
q(i) &= p(i)\sigma^2 + \sigma _a^2{\left\| \boldsymbol{\widetilde{w}}(i) \right\|^2}{v(i)^2}\\
\end{aligned}
\end{equation}
The detailed derivation is given in Appendix B. Therefore, computing the ratio between $P_K(i)$ and $P_H(i)$ yields
\begin{equation}
\begin{aligned}
\frac{P_K(i)}{P_H(i)}&=\left[ {{\left( {N \!-\! 1} \right)} \!+\! \frac{{2{\sigma ^2}\!\!\left(\! {q(i) \!+\! \sigma _a^2{\left\| \boldsymbol{\widetilde{w}}(i) \right\|^2}{v^2(i)}} \!\right)\!}}{{{p(i)q(i)}}} \!+\! \frac{{{q(i)}}}{{{p(i)^2}}}} \!\right]\!\sigma _a^2\\
&\leq\left[ {{\left( {N \!-\! 1} \right)} +4+ 1+ \frac{{\sigma _a^2{\left\| \boldsymbol{\widetilde{w}}(i) \right\|^2}{v^2(i)}}}{{{\left(\sigma^2 + \sigma _a^2{\left\| \boldsymbol{\widetilde{w}}(i) \right\|^2}\right)^2}}}} \!\right]\sigma _a^2\\
&\leq\left( {N +4+ \frac{v^2(i)}{4\sigma^2}} \!\right)\sigma _a^2
\end{aligned}
\end{equation}
For a bounded noise $v(i)$, there exists an upper bound $v_{max}$ so that
\begin{equation}
\label{PKH}
\frac{P_K(i)}{P_H(i)}\leq\left( {N +4+ \frac{v^2_{max}}{4\sigma^2}} \!\right)\sigma _a^2
\end{equation}
One can observe that the right side of Eq.({\ref{PKH}}) is independent of $v(i)$ and ${\left\| \boldsymbol{\widetilde{w}}(i) \right\|^2}$. Moreover, $P_H(i)$, $P_K(i)$ and ${\left\| \boldsymbol{\widetilde{w}}(i) \right\|^2}$ are all non-negative values. Thus combining Eq.({\ref{KH}}), Eq.({\ref{KHE}}) and Eq.({\ref{PKH}}), we can obtain
\begin{equation}
\label{CKH}
C_K(i)\leq\left( {N +4+ \frac{v^2_{max}}{4\sigma^2}} \!\right)\sigma _a^2C_H(i)
\end{equation}
Substituting Eq.({\ref{CKH}}) into Eq.({\ref{stepsize}}), we can get the following sufficient convergence condition for step size
\begin{equation}
\label{muBounded}
0<\mu<\frac{2}{\left( {N +4+ \frac{v^2_{max}}{4\sigma^2}} \!\right)\sigma _a^2}
\end{equation}

\subsection{Sufficient condition for convergence under Gaussian sensing matrix: Gaussian noise case}
Here we derive the further results of Eq.({\ref{PHiKi}}) under Gaussian noise. We assume that $v(i)$ is i.i.d Gaussian noise with zero mean and variance $\sigma_v^2$. Similarly, adopting the integral method to calculate the expectation of $P_H(i)$ and $P_K(i)$ leads to
\begin{equation}
\begin{aligned}
P_H(i)&={\frac{{\sigma\left( {\sigma^2 + \sigma _v^2} \right)\sigma_a^2}}{{{{\left( {\sigma _a^2{{\left\| {\boldsymbol{\widetilde{w}}(i)} \right\|}^2} + \sigma^2+ \sigma _v^2 } \right)}^{\frac{3}{2}}}}}}\\
P_K(i)&=(N\!-\!1)P_H(i)\sigma_a^2 + 3P_H(i){\frac{{\left( {\sigma^2 + \sigma _v^2} \right)\sigma_a^2}}{{{{ {\sigma _a^2{{\left\| {\boldsymbol{\widetilde{w}}(i)} \right\|}^2} \!+\! \sigma^2\!+\! \sigma _v^2 } }}}}}
\end{aligned}
\end{equation}
Thus one can derive the relation between $P_H(i)$ and $P_K(i)$
\begin{equation}
\begin{aligned}
P_K(i)&\leq(N-1)\sigma_a^2P_H(i)+3\sigma_a^2P_H(i)\\
&=(N+2)\sigma_a^2P_H(i)
\end{aligned}
\end{equation}
Therefore, the relation between $C_K(i)$ and $C_H(i)$ can be derived as
\begin{equation}
C_K(i)<\sigma_a^2(N+2)C_H(i)
\end{equation}
and the sufficient condition for convergence will be
\begin{equation}
\label{muGaussian}
0<\mu<\frac{2}{(N+2)\sigma_a^2}
\end{equation}
$Remark~3$: It is easy to conclude that the above condition is also applicable to zero mean Gaussian mixture noise, where the probability density function of noise is sum of several independent zero mean Gaussian processes with different variances. \\
$Remark~4$: One should note that the bounds of step size $\mu$ in Eq.(\ref{muRade}), Eq.(\ref{muBounded}) and Eq.(\ref{muGaussian}) are obtained in the extreme cases. The actual range for convergence in specific noise environment may be wider than the aforementioned bounds. In practical CS reconstruction application, one can adjust the $\mu$ based on the analysis results to achieve desirable convergence performance.

\section{Experimental Results}
In this section, we present simulation results to verify the performance of the proposed $l_0$-MCC and MB-$l_0$-MCC algorithms. The none zero entries of sparse signal $\boldsymbol{x}$ are generated from specific distribution and the corresponding positions are randomly selected and uniformly distributed within $[1,N]$. The entries of sensing matrix $\boldsymbol{\Phi}$ are independently generated from the Gaussian distribution with zero mean and variance $1/M$.
\par Two non-Gaussian noise models are used for performance verification in the simulation. The first is a two-component Gaussian mixture model(GMM) with the probability density function
\begin{equation}
p_v(i)=(1-c)N(0,\frac{\sigma_A^2}{M})+cN(0,\sigma_B^2)
\end{equation}
where $N(0,\frac{\sigma_A^2}{M})$ represents general noise disturbance with variance ${\sigma_A^2}/{M}$, and $N(0,\sigma_B^2)$ stands for outliers that occur occasionally with a large variance $\sigma_B^2$. The variable $c$ controls the occurrence probability of outliers.
\par The second is a symmetric $\alpha$-stable noise with the characteristic function
\begin{equation}
\varphi(t)=\exp(-{\gamma}^{\alpha}|t|^{\alpha})
\end{equation}
where $0<\alpha\leq2$ is the characteristic exponent. Smaller $\alpha$ yields more impulsive noise. $\gamma$ is the scale parameter. The $\alpha$-stable noise reduces to the zero mean Gaussian distribution when $\alpha=2$, while for $\alpha<2$ the distribution corresponds to an impulsive noise with infinite variance.
\par The reconstruction performance is evaluated by MSD which is approximated as the ensemble average of squared deviations
$$MSD(i)=\frac{1}{T}\sum_{t=1}^{T}{{{\left\| {{\boldsymbol{w}}^{(t)}(i)}-\boldsymbol{x} \right\|}^2}}$$
where $T$ is the number of Monte Carlo runs. Further, the reconstruction is considered successful when the MSD is less than $5\times10^{-2}$. The probability of successful reconstructions is defined as $p=T_c/T$ where $T_c$ is the amount of successful reconstructions.
\par In general a relatively small kernel width $\sigma$ can effectively handle the affections of outliers. But small $\sigma$ may decrease the convergence speed. To balance the convergence speed and reconstruction accuracy, the kernel annealing method described in \cite{ITL} is employed to select $\sigma$ in a deceasing manner across iterations:
\begin{equation}
\sigma(i)=\sigma_{max}\exp\left(-\frac{\theta i}{C}\right)+\sigma_{min}
\end{equation}
where $i$ is the current iteration number, $C$ is the maximum iteration number and $\theta$ controls the exponential decay rate. We use the estimation method in LIHT and set $\sigma_{max}$ to be $0.5(y_{(0.875)}-y_{(0.125)})-\sigma_{min}$ where $y_{(q)}$ denotes the $q$-th quantile of $\boldsymbol{y}$. $\sigma_{min}$ is set as 0.03 in all simulations.
\par In the following simulations, without explicit mention, $C$ is set to $10^4$ for $l_0$-MCC and $10^5$ for MB-$l_0$-MCC. The mini batch size $S$ for MB-$l_0$-MCC is set to $0.1M$. The error tolerance $\epsilon$, the zero attraction parameter $\beta$, the decay rate parameter $\theta$ and the step size $\mu$ are experimentally set at $10^{-4}$, $10$, $20$ and $0.2$ respectively for both algorithms.

\subsection{Convergence and reconstruction performance of \texorpdfstring{$l_0$}{l_0}-MCC and MB-\texorpdfstring{$l_0$}{l_0}-MCC  }

In this subsection we present the convergence and reconstruction performance of $l_0$-MCC and MB-$l_0$-MCC. The length $N$ of the signal $\boldsymbol{x}$ is set as 1000, and the sparsity $K=40$. The sensing matrix is employed with $M=300$. The none zero entries of $\boldsymbol{x}$ are generated from uniform distribution within $[-1,1]$. Moreover, $\boldsymbol{x}$ is normalized to a unit norm vector. The GMM noise is used as the noise model, and the variance $\sigma_A^2$ is set at $0.01$.
\par
\begin{figure}[tb]
\centering
\includegraphics[width=0.9\linewidth]{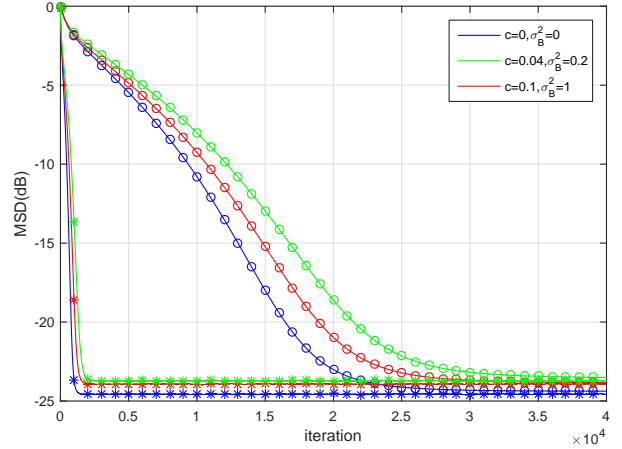}
\caption{Average learning curves of $l_0$-MCC (line with circle) and MB-$l_0$-MCC (line with star) under different noise distributions}
\label{figConvergence}
\end{figure}

 First, we examine the convergence of $l_0$-MCC and MB-$l_0$-MCC. GMM noise with different values of $c$ and $\sigma_B^2$ are tested. The regularization parameter $\lambda$ of $l_0$-MCC and $l_0$-MB-MCC are set to $5\times 10^{-6}$ and $1\times 10^{-4}$ respectively. In this simulation, 200 Monte Carlo runs with different sparse signals, sensing matrices and noises are performed. The error tolerance $\epsilon$ is not used in this simulation. The average learning curves in terms of MSD with different noise distributions are plotted in Fig.\ref{figConvergence}. Note that each weight update is considered as one iteration. One can observe that both $l_0$-MCC and MB-$l_0$-MCC can converge to the steady state with similar MSD in both Gaussian and non-Gaussian noise environment, and MB-$l_0$-MCC converges much faster than $l_0$-MCC. Moreover, the performance in heavy noise with large-value outliers($c=0.1$,$\sigma_B^2=1$) is better than that in less heavy noise with small-value outliers($c=0.04$,$\sigma_B^2=0.2$). Thus, the small outliers may have greater negative impact on steady state performance.

\begin{figure}[tb]
\centering
\includegraphics[width=0.9\linewidth]{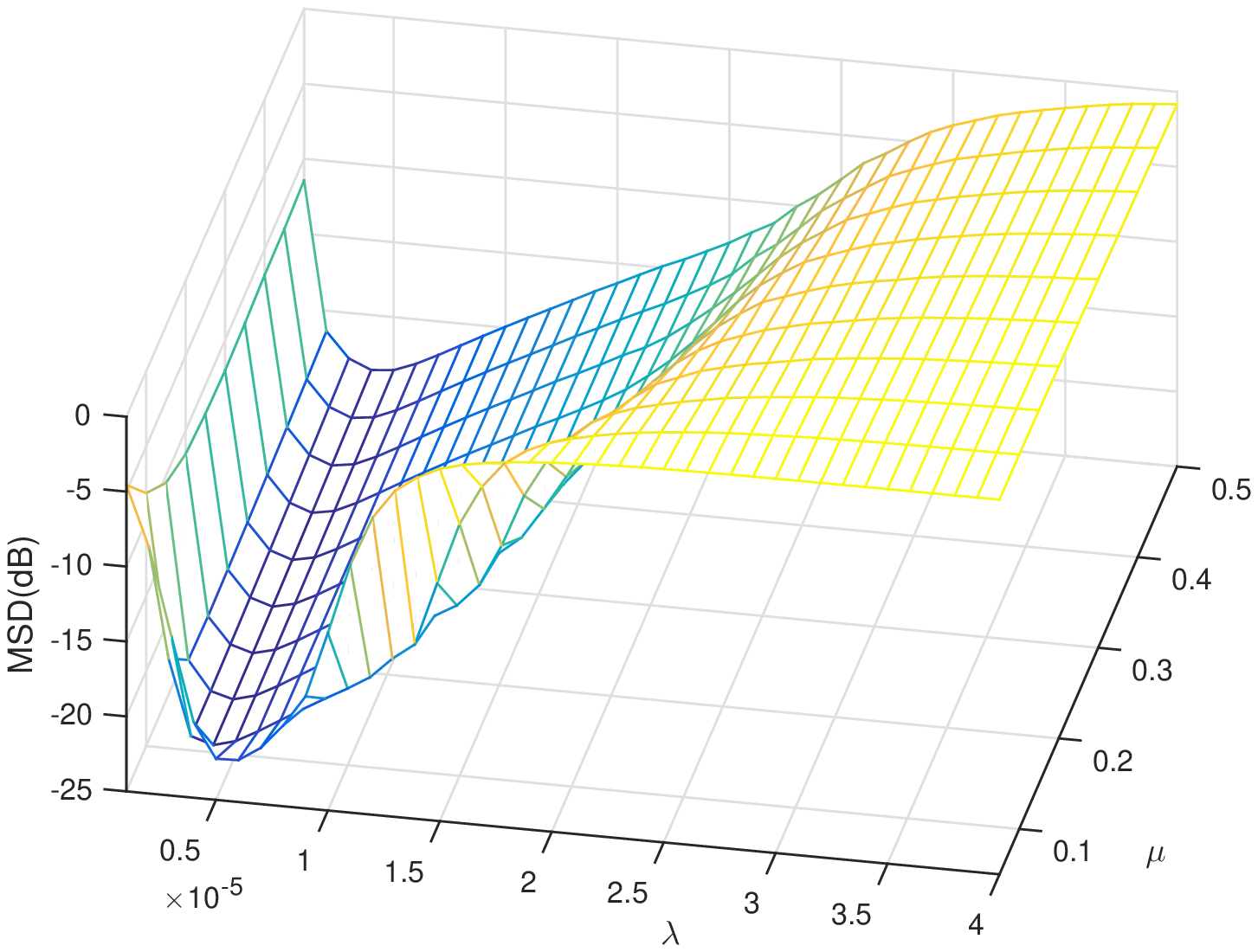}
\caption{Reconstruction MSD of $l_0$-MCC with different $\mu$ and $\lambda$}
\label{figparamG}
\end{figure}

\begin{figure}[tb]
\centering
\includegraphics[width=0.9\linewidth]{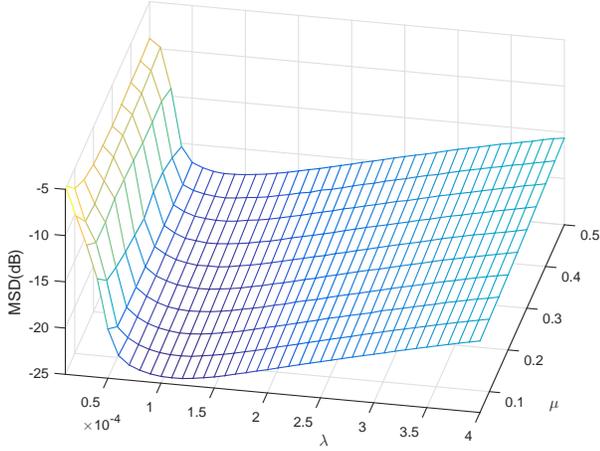}
\caption{Reconstruction MSD of MB-$l_0$-MCC with different $\mu$ and $\lambda$}
\label{figparamMB}
\end{figure}

\par Second, we investigate the effects of step size $\mu$ and regularizer $\lambda$ on reconstruction performance. The GMM noise parameters are set as $c=0.04$ and $\sigma_B^2=0.1$. We randomly generate 200 different sparse signals, sensing matrices and GMM noises. Then, the MSD values of both algorithms with different $\mu$ and $\lambda$ are computed. The range of $\mu$ is selected as $[0.05,0.5]$ for both algorithms, while $\lambda$ is set within the range $[1\times 10^{-6}, 4\times 10^{-5}]$ for $l_0$-MCC and $[1\times 10^{-5}, 4\times 10^{-4}]$ for MB-$l_0$-MCC, respectively. The MSD surfaces of $l_0$-MCC and MB-$l_0$-MCC are illustrated in Fig.\ref{figparamG} and Fig.\ref{figparamMB}, respectively. It is readily to see that the step size $\mu$ has a slight influence on MSD except a small $\mu$ is used in $l_0$-MCC. On the contrary, the performance is relatively sensitive to $\lambda$. Specifically, $l_0$-MCC is more sensitive to $\lambda$ than MB-$l_0$-MCC. In practice, given a specific CS reconstruction task, one can use cross validation to find a proper value of $\lambda$. It is easy to implement since only the sparse signal $\boldsymbol{x}$ and noise variance are needed.

\subsection{Reconstruction performance comparison: GMM noise case}
In this subsection, we compare the reconstruction performance of $l_0$-MCC and MB-$l_0$-MCC with existing $l_0$ regularized robust reconstruction algorithms in the GMM noise environment. Comparisons to $l_0$-LAD regression solved by coordinate descent($l_1$-CD) and sign-error LMS($l_0$-SLMS), $l_1$-space orthogonal matching pursuit($l_1$-OMP), Correntropy Matching Pursuit(CMP), Lorentzian-based iterative hard thresholding(LIHT) and Huber iterative hard thresholding(HIHT) are conducted. 
\par Without explicit mention, the simulation parameters are set as the same in the previous subsection, i.e. $N=1000$, $K=40$, $M=300$, and $\sigma_A^2=0.01$, $\sigma_B^2=0.1$, $c=0.04$ for GMM noise. For $l_0$-MCC and MB-$l_0$-MCC, same parameters to second simulation in Section V.A are used. While for the state-of-the-art algorithms, the parameters are tuned until the best performance is achieved under above simulation settings. Parameters of all algorithms are then fixed during the simulations.

%

\par
\begin{figure}[tb]
\centering
\includegraphics[width=0.9\linewidth]{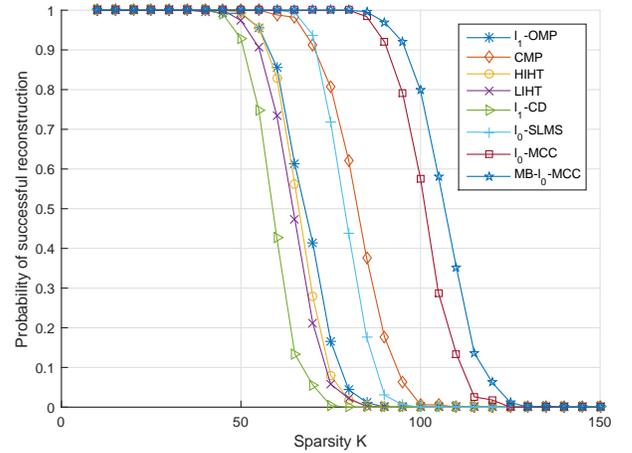}
\caption{Reconstruction probability versus sparsity K}
\label{figK}
\end{figure}

 First, we compare the reconstruction performance with different sparsities $K$. We gradually increase $K$ from 10 to 150 and calculate the reconstruction probability for all algorithms. 500 Monte Carlo runs are performed for each $K$. The reconstruction probability curves in terms of different $K$ for each algorithm are shown in Fig.\ref{figK}. One can observe that the proposed two algorithms significantly outperform other algorithms. In particular, $l_0$-MCC and MB-$l_0$-MCC can successfully reconstruct all the signals when $K$ increases up to 80, while the state-of-the-art algorithms fail when $K$ is larger than 60.

\par
\begin{figure}[tb]
\centering
\includegraphics[width=0.9\linewidth]{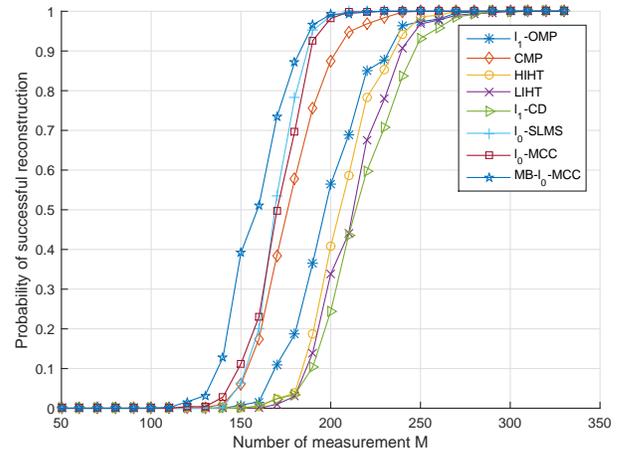}
\caption{Reconstruction probability versus number of measurements M}
\label{figM}
\end{figure}

 Second, the effect of number of measurements $M$ on reconstruction performance is investigated. We gradually increase $M$ from 50 to 330 and compute the reconstruction probability for all algorithms. For each $M$, 500 Monte Carlo runs are used to calculate the reconstruction probability. The curves of reconstruction probability versus $M$ are plotted in Fig.\ref{figM} As can be seen, $l_0$-MCC, MB-$l_0$-MCC and $l_0$-SLMS get comparable good performance and far exceed the rest algorithms. Specifically, $l_0$-SLMS can reconstruct the signal with probability one when $M$ is as small as $210$, while for $l_0$-MCC and MB-$l_0$-MCC, $M$ increases to $230$. The rest need  at least $250$ number of measurements for exact reconstruction.

\par
\begin{figure}[tb]
\centering
\includegraphics[width=0.9\linewidth]{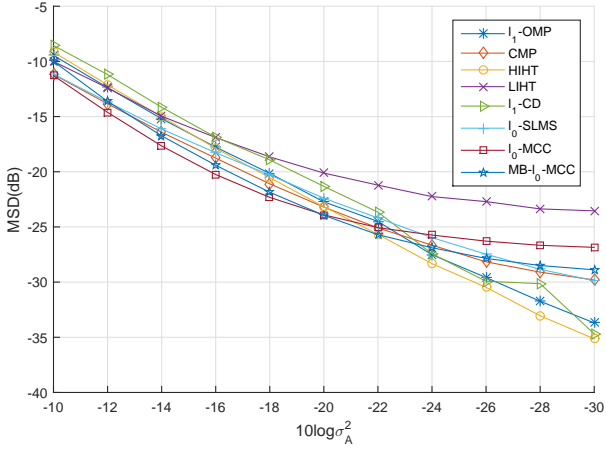}
\caption{Reconstruction MSD under different $\sigma_A^2$}
\label{figsigmaA}
\end{figure}

 Third, we explore the robustness of the algorithms to general noise. Fig.\ref{figsigmaA} illustrates the MSD of recovered signals under different values of $\sigma_A^2$. For each $\sigma_A^2$, 500 Monte Carlo runs are carried out. It shows that $l_0$-MCC and MB-$l_0$-MCC perform better than other algorithms when $\sigma_A^2$ is larger than -20dB, while when noise is small($\sigma_A^2<-24$dB) the reconstruction accuracy of the proposed algorithms is limited. Note that $l_0$-SLMS suffers from the same problem with $l_0$-MCC and MB-$l_0$-MCC. This result implies that the noise level affects the selection of regularization parameter $\lambda$, thus $\lambda$ needs to be readjusted for smaller noise to achieve lower reconstruction MSD. Nonetheless, the performance with fixed $\lambda$ is good enough if the requirement of reconstruction accuracy is not very high.

\par
\begin{figure}[tb]
\centering
\includegraphics[width=0.9\linewidth]{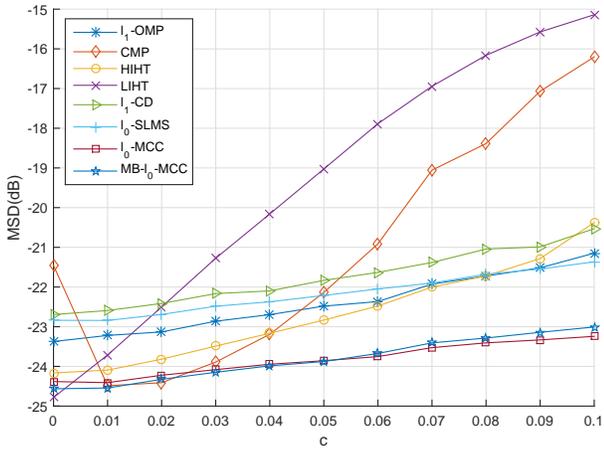}
\caption{Reconstruction normalized MSD under different $c$}
\label{figc}
\end{figure}

\begin{figure}[tb]
\centering
\includegraphics[width=0.9\linewidth]{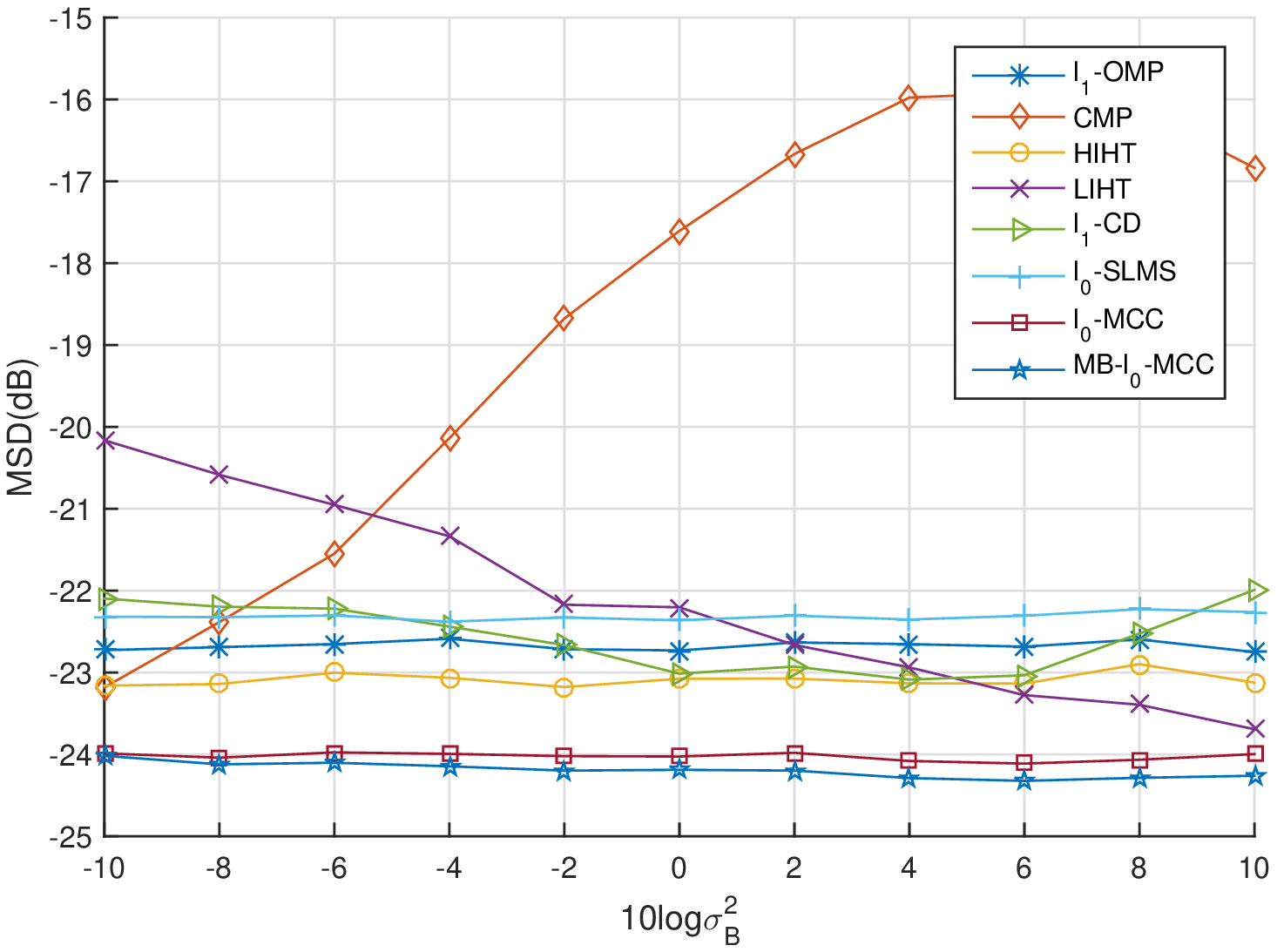}
\caption{Reconstruction MSD under different $\sigma_B^2$}
\label{figsigmaB}
\end{figure}

\begin{figure}[tb]
\centering
\includegraphics[width=0.9\linewidth]{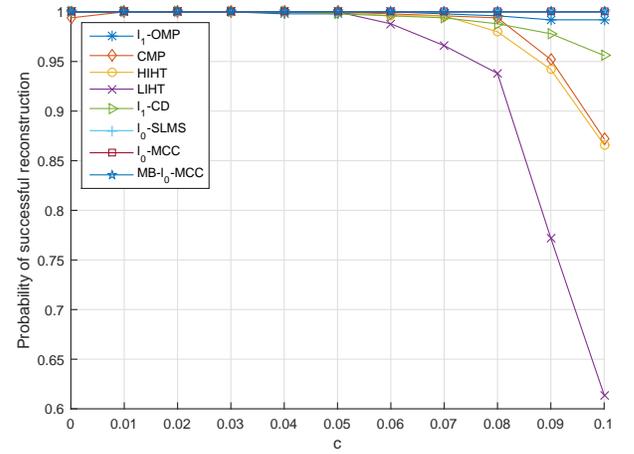}
\caption{Reconstruction probability versus $c$}
\label{figccount}
\end{figure}

 Finally, the reconstruction performance comparison under different impulsive noises is conducted. We gradually increase the outlier occurrence probability $c$ or variance $\sigma_B^2$ and compute the MSD for all algorithms. For each $c$ or $\sigma_B^2$, 500 Monte Carlo runs are conducted to calculate the reconstruction probability for each algorithm. Fig.\ref{figc} and Fig.\ref{figsigmaB} depict the MSD under different $c$ and $\sigma_B^2$, respectively. Note that only the successful reconstruction are involved in calculation of MSD.
Moreover, the curves of reconstruction probability with different $c$ are shown in Fig.\ref{figccount}. One can see that both $l_0$-MCC and MB-$l_0$-MCC achieve a lower reconstruction MSD than other algorithms. In addition, the performance of all algorithms degrades when $c$ increases, while $\sigma_B^2$ affects slightly on performance of all algorithms except for CMP and LIHT. In particular, the performance of LIHT degrades seriously when $c$ increases or $\sigma_B^2$ decreases. This may be caused by improper selection of the scale parameter $\tau$ which is fixed during the whole iterations. While CMP suffers the same problem with LIHT, which is mainly caused by the improper estimation of correntropy kernel width. Further, one can see HIHT, LIHT, $l_1$-CD, $l_1$-OMP and CMP have different degrees of reduction on reconstruction probability for heavy noises, as evident in Fig.\ref{figccount}. But adaptive filtering based algorithms(i.e. $l_0$-MCC, MB-$l_0$-MCC and $l_0$-SLMS) can still reconstruct all the signals.

\subsection{Reconstruction performance comparison: $\alpha$-stable noise case}
\begin{figure}[tb]
\centering
\includegraphics[width=0.9\linewidth]{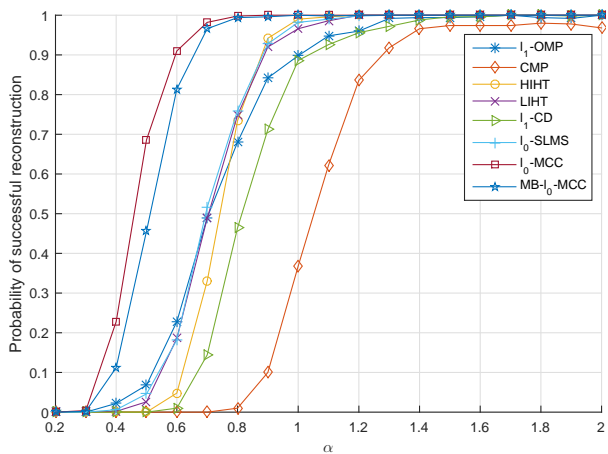}
\caption{Reconstruction probability with different $\alpha$}
\label{figalpha}
\end{figure}
\begin{figure}[tb]
\centering
\includegraphics[width=0.9\linewidth]{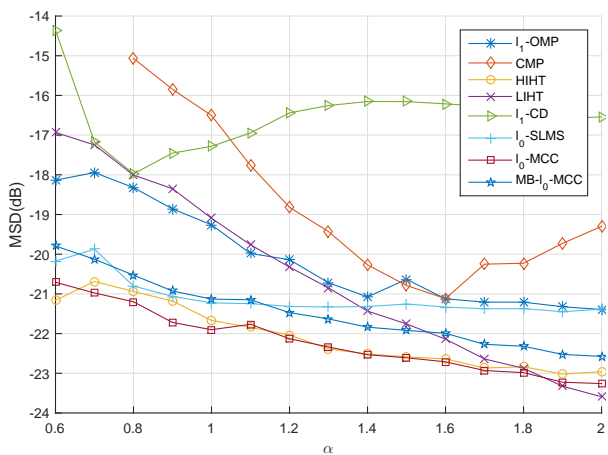}
\caption{Reconstruction MSD with different $\alpha$}
\label{figalphaMSD}
\end{figure}

In this subsection, we compare the reconstruction performance of all algorithms in $\alpha$-stable noise. The simulation parameters are set as $N=1000$, $M=300$ and $K=20$. The nonzero entries of sparse signal $\boldsymbol{x}$ are drawn from uniform distribution within range $[-1,-0.5]\bigcup[0.5,1]$ followed by a normalization. The scale parameter $\gamma$ of $\alpha$-stable noise is set to 0.01. Parameters of all algorithms are adjusted to achieve optimal reconstruction performance under $\alpha=1$. Particularly, the $\lambda$ used in $l_0$-MCC ,MB-$l_0$-MCC and $l_0$-SLMS are set to $5\times 10^{-4}$, $1.5\times 10^{-5}$ and $1\times 10^{-3}$ respectively, and the decay rate parameter $\theta$ is set to $15$ for both proposed methods. The iteration number of $l_1$-CD is set to 29.
\par For each algorithm, we gradually increase the scale parameter $\alpha$ from 0.2 to 2 and evaluate the MSD between the recovered sparse signal and true signal $\boldsymbol{x}$. The MSD is estimated over 500 Monte Carlo runs, and only successful reconstructions are involved in calculation. Fig.\ref{figalpha} depicts the reconstruction probability versus $\alpha$ for each algorithms, and the corresponding MSD curves($\alpha\geq0.6$) are shown in Fig.\ref{figalphaMSD}. As can be seen, $l_0$-MCC and MB-$l_0$-MCC achieve significantly better performance than other algorithms. In detail, both proposed algorithms get lower reconstruction MSD as well as high reconstruction probability when noise is very heavy($0.6<\alpha<1$). Moreover, $l_1$-CD experiences poor reconstruction performance when $\alpha$ is large, which is mainly caused by the fix iteration number during the simulation.


\subsection{Image Reconstruction performance comparison}
\begin{figure*}[tb]
\centering
\includegraphics[width=0.99\linewidth]{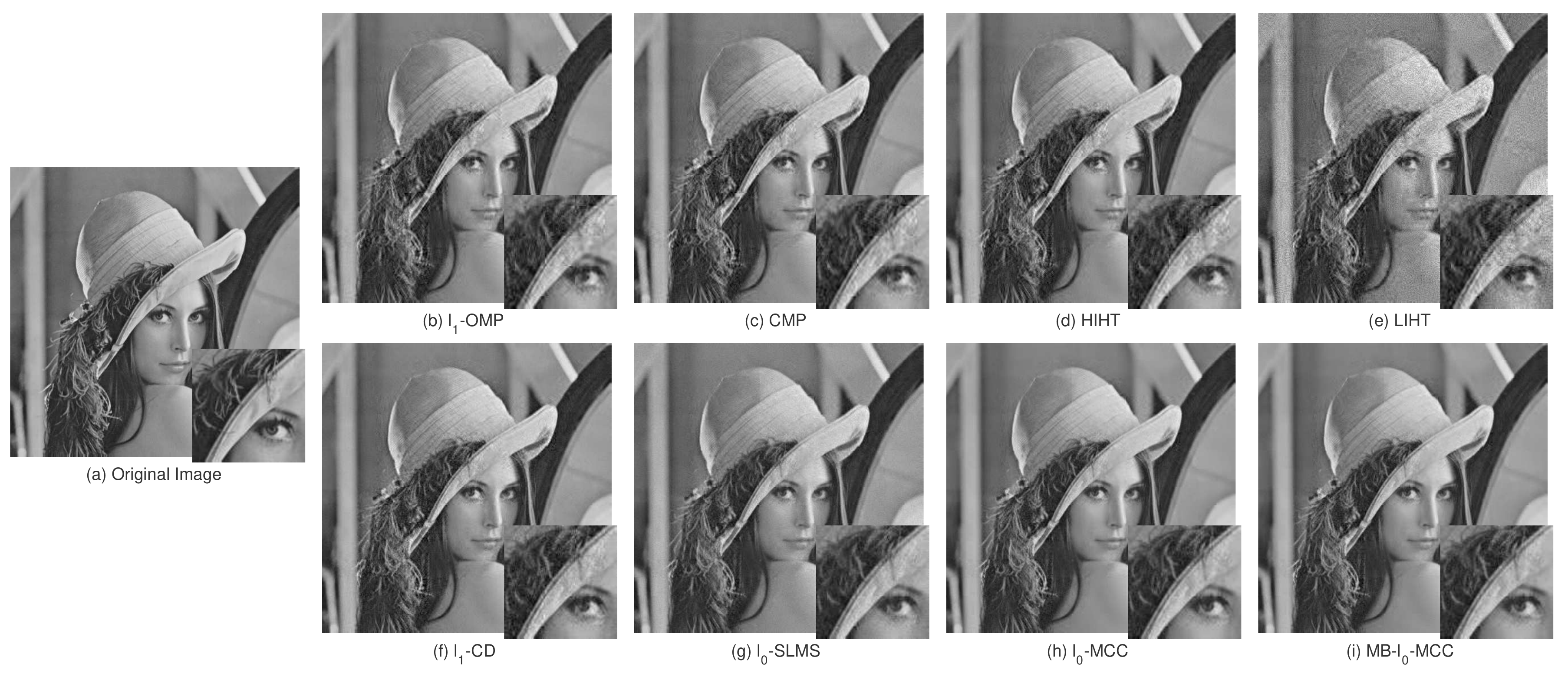}
\caption{Image reconstruction example of lena image under GMM noise. No sparsification is performed on the transformed DCT signals. (a) Original image. (b)-(i) Reconstructed images using $l_1$-OMP(PSNR=28.80dB), CMP(PSNR=29.25dB), HIHT(PSNR=28.98dB), LIHT(PSNR=24.75dB), $l_1$-CD(PSNR=29.90dB), $l_0$-SLMS(PSNR=30.10dB), $l_0$-MCC(PSNR=31.91dB) and MB-$l_0$-MCC(PSNR=31.74dB), respectively. The bottom-right of each image is the partial enlarged view.
}
\label{figimagesample}
\end{figure*}

In this section we compare the reconstruction performance on natural images. We choose Lena, Barbara, Cameraman and Peppers as the test images. All the test images are $512\times 512$. We use block-based compressive sensing techniques\cite{Gan2007Block}. The image is separated into small patches (in this simulation the patch size is set as $32\times 32$). The 2D DCT transformation is then used to get sparse representation. For each transformed signal, the sparsity is set as $s$, i.e. the largest $s$ coefficients are selected to approximate the signal. The measurement number $M$ is set as 500 so that a $500\times 1024$ Gaussian random matrix is used as the sensing matrix. To evaluated the reconstruction performance in noisy measurements, the Gaussian and GMM noise are added to the clean measurements. For GMM noise, the parameters are set as $\sigma_A^2=0.04$, $\sigma_B^2=10$, $c=0.02$. While for Gaussian noise, the variance is set as $0.04/M$. For $l_0$-MCC, MB-$l_0$-MCC and $l_0$-SLMS, the normalized noisy measurements are used to perform the reconstruction. The recovered signals are then multiplied by the normalized factor to obtain final sparse estimation. Parameters of all algorithms are tuned to achieve best performance. The algorithms are implemented in MATLAB with two 2.6GHz processors and 64 GB memory. The block processing is implemented using parallel method with 20 cores. Table \ref{imagecomparison} shows the reconstruction PSNR and average running times for each algorithm in different sparsity and noise situations. $s=1024$ denotes that no sparsification is performed on transformed DCT signals, and in this case the sparsity parameter is experimentally set to 50 for $l_1$-OMP, CMP, LIHT and HIHT. From the results one can see that when $s=100$, $l_1$-OMP, $l_1$-CD, $l_0$-MCC and MB-$l_0$-MCC can achieve comparable high values of PSNR in both Gaussian and non-Gaussian situations, while $l_0$-MCC and MB-$l_0$-MCC runs much faster than $l_1$-CD and $l_1$-OMP. When $s=1024$, the signals may be non-sparse or nearly sparse. In this case the reconstruction is more challenging. As can be seen, the performance of all the algorithms degrades when $s=1024$, while $l_0$-MCC and MB-$l_0$-MCC can still achieve good performance and outperform other algorithms significantly. Fig.\ref{figimagesample} shows the reconstructed Lena images of all algorithms under $s=1024$ and GMM noise. It can be seen that the reconstructed images using $l_0$-MCC and MB-$l_0$-MCC are more accurate than others.\\

\renewcommand{\arraystretch}{1.5} 
\newcolumntype{C}[1]{>{\centering\arraybackslash}p{#1}}
\begin{table*}[tp]
  \centering
  \fontsize{6.5}{8}\selectfont
  \begin{threeparttable}
  \caption{Image Reconstruction Performance Comparison: Reconstruction PSNR and average running times}
  \label{imagecomparison}
    \begin{tabular}{C{1.25cm}C{0.7cm}C{0.6cm}C{0.65cm}C{0.7cm}C{0.6cm}C{0.65cm}C{0.7cm}C{0.6cm}C{0.65cm}C{0.7cm}C{0.6cm}C{0.65cm}C{0.7cm}C{0.6cm}C{0.65cm}}
    \toprule
    \multirow{3}{*}{Method}&
    \multicolumn{3}{c}{ Lena}&\multicolumn{3}{c}{ Barbara}&\multicolumn{3}{c}{ Cameraman}&\multicolumn{3}{c}{ Peppers}&\multicolumn{3}{c}{Average running time(s)}\cr
    \cmidrule(lr){2-4} \cmidrule(lr){5-7} \cmidrule(lr){8-10} \cmidrule(lr){11-13} \cmidrule(lr){14-16}
    &Gaussian&GMM&GMM&Gaussian&GMM&GMM&Gaussian&GMM&GMM&Gaussian&GMM&GMM&Gaussian&GMM&GMM\cr
    &$s$=100&$s$=100&$s$=1024&$s$=100&$s$=100&$s$=1024&$s$=100&$s$=100&$s$=1024&$s$=100&$s$=100&$s$=1024&$s$=100&$s$=100&$s$=1024\cr
    \midrule
    $l_1$-OMP&33.57& 33.23& 28.80& 31.07& 30.77& 25.98& 35.18& 34.80& 29.48& 32.77&{\bf 32.51}& 28.81&840.00&853.82&326.79\cr
    CMP&31.36& 31.58& 29.25& 29.14& 30.27& 26.52& 33.10& 32.62& 29.86& 31.56& 30.70& 29.10&146.52&65.48&30.19\cr
    HIHT&33.90& 32.45& 28.98& 31.58& 29.49& 26.18& {\bf 35.63}& 33.69& 29.90& 32.93& 31.90& 29.03&6.35&7.74&3.88\cr
    LIHT&{\bf 33.97}& 19.34& 24.75& {\bf 31.60}& 19.80& 24.01&{\bf 35.63}& 19.88& 25.01& {\bf 33.05}& 20.01& 25.14&7.77&10.72&5.51\cr
    $l_1$-CD&33.66& 33.27& 29.90& 31.19& 30.61& 27.11& 35.27& 34.77& 31.33& 32.74& 32.42& 29.74&191.08&187.65&189.82\cr
    $l_1$-SLMS&31.58& 30.87& 30.10& 29.64& 29.02& 28.06& 32.81& 31.80& 31.00& 31.05& 30.35& 30.07&17.46&17.08&16.72\cr
    $l_0$-MCC&33.61&{\bf 33.34}&{\bf 31.91}& 31.12&{\bf 30.87}&{\bf 29.18}& 35.24& 34.80&{\bf 33.58}& 32.71&32.39&{\bf 31.48}&17.45&17.12&17.35\cr
    MB-$l_0$-MCC&33.70& 33.31& 31.74& 31.19& 30.85& 29.05& 35.35&{\bf 34.81}& 33.31& 32.79& 32.38& 31.33&10.76&10.72&10.82\cr
    \bottomrule
    \end{tabular}
    \end{threeparttable}
\end{table*}

\section{Conclusion}
In this paper we proposed two gradient based robust CS reconstruction algorithms, namely, $l_0$-MCC and MB-$l_0$-MCC. By using adaptive filtering framework and adding a zero attraction term, the proposed algorithms can adaptively estimate the sparsity and accurately recover the signals. In particular, by taking the advantages of correntropy, the new algorithms can robustly recover the sparse signal in non-Gaussian noise situations. Moreover, the mini batch approach can significantly improve the convergence speed. Theoretical analysis gives some sufficient conditions to ensure the convergence of $l_0$-MCC. Simulation results illustrate that the proposed algorithms can achieve better performance than existing robust algorithms under different non-Gaussian noise environments. Simulations also show that the performance of the new algorithms heavily depends on the regularizer $\lambda$. For a specific CS application, we need to choose a proper value of $\lambda$ to get desirable performance. How to optimize the regularizer in an online manner is an interesting topic for future study.
\section*{Acknowledgements}
This work was supported by National Natural Science Foundation of China under Grant No.91648208 and No.61231018, program of introducing talents of discipline
to university of China under Grant B13043 and National Science and Technology support program of China under Grant 2015BAH31F01.

\begin{appendices}
\section{Bound of $B(i)$ in Eq.(\ref{Bii})}
We first define $R(i)$ as
\begin{equation}
R(i)=E[G(e(i))v^2(i)]\\
\end{equation}
It holds that
\begin{equation}
\begin{aligned}
0\leq R(i)\leq E[v^2(i)]
\end{aligned}
\end{equation}
Thus $R(i)$ is bounded based on the finite noise variance assumption. Therefore
\begin{equation}
\begin{aligned}
\!\left| {J( i)}\!+\!R(i) \right| &\!=\! \left| {E\!\left[ {G(e(i))v(i){{\boldsymbol{\phi}}^T}(i)\boldsymbol{\widetilde{w}}(i)} \right]}\!\!+\!E[G(e(i))v^2(i)] \right|\\
&\!= \left| {E\left[ {G(e(i))e(i)v(i)} \right]} \right|\\
&\!\leq  {E\left[ \left|G(e(i))e(i)\right|\left|v(i)\right| \right]} \\
&\!\mathop \le \limits^{\left( a \right)}  \frac{\sqrt{2}\sigma}{e}{E\left[ \left|v(i)\right| \right]} \\
\end{aligned}
\end{equation}
where $(a)$ comes from the fact that
\begin{equation}
0\leq\exp(-\frac{x^2}{2\sigma^2})x\leq\frac{\sqrt{2}\sigma}{e}\\
\end{equation}
So $J(i)$ is bounded by
\begin{equation}
\begin{aligned}
-\frac{\sqrt{2}\sigma}{e}{E\left[ \left|v(i)\right| \right]}-E[v^2(i)]\leq J(i)\leq \frac{\sqrt{2}\sigma}{e}{E\left[ \left|v(i)\right| \right]}
\end{aligned}
\end{equation}
Further, one can obtain the following bound of $T(i)$:
\begin{equation}
\begin{aligned}
 {T\left( i \right)} &= E\left[G^2(e(i))v^2(i)\|\boldsymbol{\phi}(i)\|^2\right]\\
 &\le E\left[ {{v^2}(i){{\left\| {\boldsymbol{\phi}(i)} \right\|}^2}} \right]\\
 &= N\sigma _a^2E\left[ {{v^2}\left( i \right)} \right]
\end{aligned}
\end{equation}
In addition, we can derive
\begin{equation}
\begin{aligned}
\left| {Y(i)}+{T\left( i \right)} \right| &= \left| {E\left[ {{G^2}\left( {e\left( i \right)} \right)v\left( i \right){{\left\| {\boldsymbol{\phi}\left( i \right)} \right\|}^2}{\boldsymbol{\phi}^T}\left( i \right)\boldsymbol{\widetilde{w}}\left( i \right)} \right]}\right.\\
&~~~~\left.+E\left[G^2(e(i))v^2(i)\|\boldsymbol{\phi}(i)\|^2\right] \right|\\
&=\left|E\left[G^2(e(i))e(i)v(i)\|\boldsymbol{\phi}(i)\|^2\right]\right|\\
&\leq E\left[\left|G^2(e(i))e(i)\right|\left|v(i)\right|\|\boldsymbol{\phi}(i)\|^2\right]\\
&\leq \frac{\sigma}{e}N\sigma _a^2E\left[\left|v(i)\right|\right]
\end{aligned}
\end{equation}
Thus, $Y(i)$ is bounded by
\begin{equation}
-E\left[\frac{\sigma}{e}\left|v(i)\right|+ {{v^2}\left( i \right)} \right]N\sigma _a^2\leq Y(i)\leq \frac{\sigma}{e}N\sigma _a^2E\left[\left|v(i)\right|\right]
\end{equation}
For $P(i)$, we can get
\begin{equation}
\begin{aligned}
\left| {P\left( i \right)} \right| &= \left|E\left[G(e(i))e(i)(\boldsymbol{\phi}^T(i)\boldsymbol{z}_{\beta}(\boldsymbol{w}(i)))\right] \right|\\
 &\le E\left[\left|G(e(i))e(i)\right|\left|\boldsymbol{\phi}^T(i)\boldsymbol{z}_{\beta}(\boldsymbol{w}(i))\right|\right] \\
 &\mathop \le \limits^{\left( a \right)} \sum\limits_{k = 1}^N {E\left[ {\frac{\sigma }{e}\left| {{{\phi}_k}\left( i \right)} \right|\left| {{z_\beta }\left( {{{\tilde w}_k}\left( i \right)} \right)} \right|} \right]} \\
 &< \sum\limits_{k = 1}^N {E\left[ {\frac{\sigma }{e}\left| {{{\phi}_k}\left( i \right)} \right|\beta } \right]} \\
 &= N\frac{{\sigma \beta }}{e}E[\left| {{{\phi}_1}\left( i \right)} \right|]
\end{aligned}
\end{equation}
Moreover, $M(i)$ and $Q(i)$ have been proved to be bounded with\cite{Jin2010A}
\begin{equation}
\begin{aligned}
|M(i)|&\leq N+\beta\|\boldsymbol{x}\|_1\\
Q(i)&\leq N\beta^2
\end{aligned}
\end{equation}
Finally, all the terms on the right side of Eq.(\ref{Bii}) are bounded, thus $B(i)$ is bounded.

\section{Derivation of $P_K(i)$ and $P_H(i)$ in Eq.(\ref{PHiKi})}
Since the elements of ${\boldsymbol{\phi}}(i)$ are independent and jointly Gaussian with variance $\sigma_a^2$, ${{\delta}}(i)= {{\boldsymbol{\phi}}^T(i)}{\boldsymbol{h}}$ will also be Gaussian with
\begin{equation}
E\left( {{\delta}}(i) \right) = \sum_{k=1}^{N} h_k E\left( {\phi_k(i)} \right)= 0
\end{equation}
\begin{equation}
\begin{aligned}
E\left[ {{{{\delta}}^2(i)}} \right] = \sum_{k=1}^{N} h_k^2 E\left( {\phi_k^2(i)} \right)= \sigma _a^2{\left\| {\boldsymbol{h}} \right\|^2}
\end{aligned}
\end{equation}
that is, $\delta  \sim N\left( {0,\sigma _a^2{{\left\| {\boldsymbol{h}} \right\|}^2}} \right)$. Using integral method to compute the expectation one can obtain
\begin{equation}
\begin{aligned}
&f(\varepsilon,{\boldsymbol{h}})\\
&=\int\limits_{ - \infty }^{ + \infty } {\exp \left( { - \frac{{{{\left( {\varepsilon - {{\boldsymbol{\phi}}^T}(i){\boldsymbol{h}}} \right)}^2}}}{{2{\sigma ^2}}}} \right){{\left( {{{\boldsymbol{\phi}}^T}(i){\boldsymbol{h}}} \right)}^2}} d{\boldsymbol{\phi}}\\
&\!= \int\limits_{ - \infty }^{ + \infty } { {\left[ {\exp \left( { - \frac{{{{\left( {\varepsilon - \delta } \right)}^2}}}{{2{\sigma ^2}}}} \right){\delta ^2}\left( {\frac{\exp \left( { - \frac{{{\delta ^2}}}{{2\sigma _a^2{{\left\| {\boldsymbol{h}} \right\|}^2}}}} \right)}{{\sqrt {2\pi \sigma _a^2{{\left\| {\boldsymbol{h}} \right\|}^2}} }}} \right)} \right]d\delta } }\\
&\!= \!{\frac{{\sigma \!\!\left( {{\sigma ^4}\!+\!{{\left\| {\boldsymbol{h}} \right\|}^2}\sigma _a^2({\sigma ^2}\!+\!{\varepsilon^2})}\! \right)\!}}{{{{\left( {\sigma _a^2{{\left\| {\boldsymbol{h}} \right\|}^2} + {\sigma ^2}} \right)}^{\frac{5}{2}}}}}}{\sigma _a^2{{\left\| {\boldsymbol{h}} \right\|}^2}\!\exp \!\left(\! { - \frac{{{\varepsilon^2}}}{{2\!\left(\!  {\sigma _a^2{{\left\| {\boldsymbol{h}} \right\|}^2\!}\! +\! {\sigma ^2}}  \right)}}} \!\right)}
\end{aligned}
\end{equation}
\begin{equation}
\begin{aligned}
&g_k(\varepsilon,{\boldsymbol{h}})\\
&=\int\limits_{ - \infty }^{ + \infty } {\exp \left( { - \frac{{{{\left( {\varepsilon - {{\boldsymbol{\phi}}^T}(i){\boldsymbol{h}}} \right)}^2}}}{{2{\sigma ^2}}}} \right){{\left( {{{\boldsymbol{\phi}}^T}(i){\boldsymbol{h}}} \right)}^2}} \phi^2_k(i)d{\boldsymbol{\phi}}\\
&=\int_{ - \infty }^{ + \infty } \left[ \exp \left( { - \frac{{{{\left( {\varepsilon - {{\tilde \delta }_k} - {h_k}{\phi_k}} \right)}^2}}}{{2{\sigma ^2}}}} \right){{\left( {{{\tilde \delta }_k} + {h_k}{\phi_k}} \right)}^2}\right.\\
&~~~~~~~~~~~~~~\left. \times \phi_k^2\frac{{\exp \left( { - \frac{{\tilde \delta _k^2}}{{2\sigma _a^2\left( {{h^2} - h_k^2} \right)}} - \frac{{u_k^2}}{{2\sigma _a^2}}} \right)}}{{2\pi \sigma _i^2\sqrt {{h^2} - h_k^2} }} \right]d{{\tilde \delta }_k}d{\phi_k}\\
& = \left[\frac{{\sigma \!\left( \!{{\sigma ^2}\!+\!\sigma _a^2\left( {{{\left\| {\boldsymbol{h}} \right\|}^2} \!-\! h_k^2} \right)} \!\right)\!\left(\! {{\sigma ^4}\!+\!{{\left\| {\boldsymbol{h}} \right\|}^2}\sigma _a^2\left( {{\sigma ^2} \!+\! {\varepsilon^2}} \right)} \!\right)\!{{\left\| {\boldsymbol{h}} \right\|}^2}}}{{{{\left( {{\sigma ^2}+\sigma _a^2{{\left\| {\boldsymbol{h}} \right\|}^2}} \right)}^{\frac{7}{2}}}}}\right.\\
&~~~~~ + \frac{{\sigma {{\left( {{\sigma ^4}+{{\left\| {\boldsymbol{h}} \right\|}^2}\sigma _a^2\left( {{\sigma ^2}+{\varepsilon^2}} \right)} \right)}^2}h_k^2}}{{{{\left( {{\sigma ^2}+\sigma _i^2{{\left\| {\boldsymbol{h}} \right\|}^2}} \right)}^{\frac{9}{2}}}}}\\
&~~~~~ \left.+ \!\frac{{{\sigma ^3}\!\left(\! {{\sigma ^4}\!\!+\!\!{{\left\| {\boldsymbol{h}} \right\|}^{\!2}}\sigma _a^2\!\left( {{\sigma ^2} \!\!+\!\! 3{\varepsilon^2}} \!\right)\!} \right)\!h_k^2}}{{{{\left( {{\sigma ^2}+\sigma _a^2{{\left\| {\boldsymbol{h}} \right\|}^2}} \right)}^{\frac{7}{2}}}}}\right]\!\exp \!\left(\! { \!- \frac{{{\varepsilon^2}}}{{2\!\left(\!  {\sigma _a^2{{\left\| {\boldsymbol{h}} \right\|}^2\!}\! +\! {\sigma ^2}}  \right)}}} \!\right)
\end{aligned}
\end{equation}
where some calculations of integral are based on the following formulas
\begin{equation}
\begin{aligned}
&\int_{{ { - }}\infty }^{{ { + }}\infty } {x^{2n{ { + }}1}}{e^{ - a{x^2}}}dx = 0 \\
&\int_{{ { - }}\infty }^{{ { + }}\infty } {x^{2n}}{e^{ - a{x^2}}}dx = 2\sqrt \pi  \frac{{\left( {2n} \right)!}}{{n!}} {\left( {\frac{1}{{4a}}} \right)^{\frac{{2n + 1}}{2}}}
\end{aligned}
\end{equation}
Thus
\begin{equation}
\begin{aligned}
H(i)&=E[f(e(i))(\boldsymbol{\phi}^T(i)\boldsymbol{\widetilde{w}}(i))^2]\\
&=E[f(v(i),\boldsymbol{\widetilde{w}}(i))]\\
&=E\left[ P_H(i){\left\| \boldsymbol{\widetilde{w}}(i) \right\|^2}\right]\\
K(i)&=E[G^2(e(i))\|\boldsymbol{\phi}(i)\|^2(\boldsymbol{\phi}^T(i)\boldsymbol{\widetilde{w}}(i))^2]\\
&\leq E[G(e(i))\|\boldsymbol{\phi}(i)\|^2(\boldsymbol{\phi}^T(i)\boldsymbol{\widetilde{w}}(i))^2]\\
&=\sum_{k=1}^{N}E[g_k(v(i),\boldsymbol{\widetilde{w}}(i))]\\
&=E\left[P_K(i){\left\| \boldsymbol{\widetilde{w}}(i) \right\|^2}\right]
\end{aligned}
\end{equation}
where $P_K(i)$ and $P_H(i)$ are shown in Eq.(\ref{PHiKi}).
\end{appendices}

\bibliographystyle{IEEEtran}
\bibliography{SMCC_CS}

\end{document}